\documentclass[a4paper,UKenglish,cleveref, autoref, thm-restate]{lipics-v2021}

\pdfoutput=1 
\hideLIPIcs  

\title{ReqGenX: An Empirical Study of Atomic Decomposition, Artifact Regeneration, and Reconstruction for Legacy SRS Documents} 

\titlerunning{ReqGenX} 

\author{Ragib Shahariar Ayon}{Texas State University, San Marcos, TX, USA \and \url{https://www.ragibayon.com/}}{ipd21@txstate.edu}{https://orcid.org/0009-0006-6372-5000}{}

\author{Rayed Fahmi}{Texas State University, San Marcos, TX, USA}{uoo22@txstate.edu}{}{}

\author{Sumon Biswas}{Case Western Reserve University,  Cleveland, OH, USA \and \url{https://sumonbis.github.io/}}{sumon@case.edu}{https://orcid.org/0000-0001-7074-1953}{}

\author{Shibbir Ahmed}{Texas State University, San Marcos, TX, USA \and \url{https://shibbirtanvin.github.io/}}{shibbir@txstate.edu}{https://orcid.org/0000-0003-1183-883X}{}

\authorrunning{R. Ayon, R. Fahmi, S. Biswas and S. Ahmed} 

\Copyright{Jane Open Access and Joan R. Public} 



\category{} 

\relatedversion{} 

\supplement{https://doi.org/10.5281/zenodo.20267334}

\nolinenumbers 

\EventEditors{John Q. Open and Joan R. Access}
\EventNoEds{2}
\EventLongTitle{42nd Conference on Very Important Topics (CVIT 2016)}
\EventShortTitle{CVIT 2016}
\EventAcronym{CVIT}
\EventYear{2016}
\EventDate{December 24--27, 2016}
\EventLocation{Little Whinging, United Kingdom}
\EventLogo{}
\SeriesVolume{42}
\ArticleNo{23}

\usepackage[utf8]{inputenc}
\usepackage[table]{xcolor}
\usepackage{xspace}
\usepackage{multirow}
\usepackage{balance} 
\usepackage{rotating}
\usepackage{listings}
\usepackage{lscape}
\usepackage[ruled,vlined]{algorithm2e}
\usepackage{tabularx} 
\usepackage{wrapfig} 
\usepackage{xurl} 
\usepackage{flushend} 
\usepackage{tikz} 
\usepackage[normalem]{ulem} 
\usepackage[most]{tcolorbox} 
\usepackage{syntax} 
\usepackage{booktabs} 
\usepackage{graphicx} 
\usepackage{caption} 
\usepackage{hyperref} 
\usepackage{framed} 
\usepackage{subcaption}
\usepackage{listings}

\usepackage{array}

\newcolumntype{L}{>{\raggedright\arraybackslash}X} 
\newcolumntype{C}{>{\centering\arraybackslash}X}   
\newcolumntype{R}{>{\raggedleft\arraybackslash}X}  

\definecolor{codegreen}{rgb}{0,0.6,0}
\definecolor{codegray}{rgb}{0.5,0.5,0.5}
\definecolor{codepurple}{rgb}{0.58,0,0.82}
\definecolor{backcolour}{rgb}{0.95,0.95,0.92}
\definecolor{medium-blue}{rgb}{0,0,1}
\definecolor{darkblue}{HTML}{003566}
\definecolor{lightgray}{gray}{0.97}

\hypersetup{
  colorlinks=true,
  linkcolor=black,
  urlcolor=medium-blue,
  citecolor=black
}

\lstdefinestyle{custompython}{
  backgroundcolor=\color{backcolour},
  commentstyle=\color{codegreen},
  keywordstyle=\color{magenta},
  numberstyle=\tiny\color{codegray},
  stringstyle=\color{codepurple},
  basicstyle=\scriptsize\ttfamily,
  breaklines=true,
  captionpos=b,
  numbers=left,
  numbersep=5pt,
  showstringspaces=false,
  tabsize=2
}
\newcommand{\name}{\textit{ReqGenX}\xspace}

\definecolor{traceReq}{HTML}{FFCCE6}
\definecolor{traceCtx}{HTML}{CCCCFF}
\definecolor{traceOmit}{HTML}{D5E8D4}
\definecolor{traceNorm}{HTML}{FFE6CC}

\usepackage{caption} 
\usepackage{caption} 
\newcounter{NumObservations}
\newcommand{\finding}[2][]{%
  \begin{tcolorbox}[
      enhanced,
      breakable,
      colback=gray!5,
      colframe=black,
      boxrule=0.5pt,
      arc=2pt,
      left=6pt,
      right=6pt,
      top=4pt,
      bottom=4pt,
      width=\linewidth
  ]
  \textbf{#1\ifx&#1&Finding~\arabic{NumObservations}\fi:}~#2
  \end{tcolorbox}
  \stepcounter{NumObservations}%
}

\usetikzlibrary{positioning,shapes.geometric,shapes.misc}
\newcommand*\circleB[1]{%
  \tikz[baseline=(char.base)]{
    \node[
      shape=circle,
      fill=gray!15!white,
      inner sep=1pt
    ] (char) {\textcolor{black}{#1}};
  }%
}

\newtcolorbox{examplebox}[1][]{
  enhanced,
  breakable,
  colback=gray!5,
  colframe=gray!50,
  boxrule=0.4pt,
  arc=2pt,
  left=6pt,
  right=6pt,
  top=4pt,
  bottom=4pt,
  width=\linewidth,
  fontupper=\small,
  title={#1},
  fonttitle=\bfseries,
  coltitle=black,
  halign title=center
}

\newcommand{\legendbox}[1]{\textcolor{#1}{\rule{1.5ex}{1.5ex}}}

\begin{document}

\maketitle

\begin{abstract}
\textbf{Background:} Evaluating automated Software Requirements Specification (SRS) generation is challenging because few datasets provide fine-grained traceability between source requirements, intermediate elicitation artifacts, and generated specifications. \textbf{Aims:} We aim to study whether legacy SRS documents can be transformed into traceable synthetic pre-SRS artifacts that support fine-grained evaluation of LLM-based SRS generation. \textbf{Method:} We conduct an empirical study using \name, a controlled pipeline that decomposes SRS sections into source-grounded atomic statements, routes atoms to standards-inspired artifact types through multi-LLM plurality voting, and generates artifacts using constrained prompts with iterative judge-guided refinement. We evaluate \name\ on seven PURE SRS documents using grounding, quality, information retention, and downstream reconstruction analyses. \textbf{Results:} \name\ produces faithful and usable atoms, with median AlignScore values typically between 0.96 and 0.99 and Prometheus scores ranging from 4.34 to 4.85. Generated artifacts remain strongly grounded in their source atoms, with AlignScore values typically between 0.80--0.94 and judge pass rates near 100\%; stricter Prometheus evaluation yields pass rates from 54.8\% to 97.1\%. In a downstream SRS reconstruction case study, artifact-backed atoms remain recoverable from generated SRSs, with SBERT means between 0.69 and 0.75 and AlignScore medians between 0.76 and 0.84. \textbf{Conclusions:} Traceable synthetic pre-SRS artifacts can support more fine-grained evaluation of LLM-based SRS generation, while exposing tradeoffs among faithfulness, information retention, and artifact completeness.
\end{abstract}

\section{Introduction}
\label{introduction}

\begin{figure}[t]
\centering
\includegraphics[width=1\textwidth]{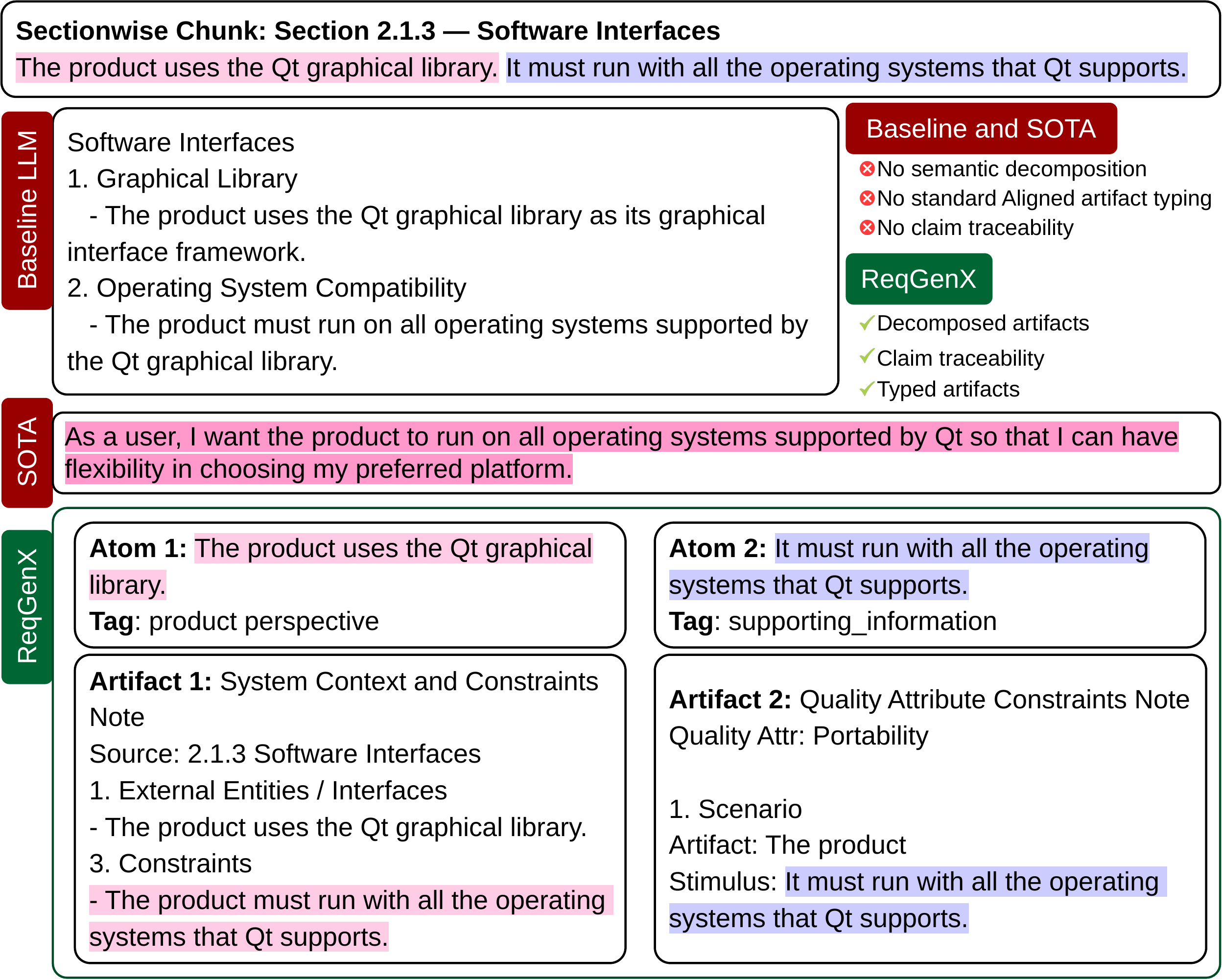}
\caption{Example of section-to-atom decomposition and artifact routing. A mixed section is split into atomic statements and mapped to different artifact types. The baseline merges claims into a single artifact, whereas \name produces typed, traceable artifacts. Colored highlights show mappings from source statements to artifacts (\legendbox{traceReq}: requirement content, \legendbox{traceCtx}: supporting information content).}
\label{fig:chunk-two-artifacts}
\end{figure}

Large language models (LLMs) are increasingly being applied to requirements engineering (RE), including SRS generation, requirement extraction, validation, correction, and the production of intermediate artifacts such as user stories and acceptance criteria~\cite{krishna2024using,zhu2025reqinone,spijkman2025llm}. However, SRS compilation differs from many software engineering tasks because requirements are primarily expressed in natural language and encode stakeholder intent, system scope, constraints, assumptions, and quality expectations. As a result, a generated SRS may appear coherent while omitting source information, merging distinct concerns, or introducing unsupported requirements. Evaluating LLM-based SRS compilation, therefore, requires more than document-level fluency: it requires claim-level evidence about which source statements support each generated artifact and whether the original requirements knowledge is preserved. Prior Natural Language Processing for Requirements Engineering (NLP4RE) work provides important foundations for this problem, including requirements classification, extraction, defect detection, quality analysis, and traceability~\cite{zhao2021natural}. Public corpora such as PURE provide requirements documents for experimentation~\cite{ferrari2017pure}, and traceability research shows that intermediate artifacts can help bridge abstraction gaps among lifecycle artifacts~\cite{gao2024triad}. However, existing resources do not provide a ready-made evaluation substrate that jointly links heterogeneous SRS evidence, intermediate RE artifacts, generated requirements, and traceability relations. This motivates using legacy SRSs as seed evidence for constructing traceable synthetic RE resources that can support fine-grained evaluation of LLM-based SRS compilation.

Figure~\ref{fig:chunk-two-artifacts} illustrates this challenge using a \textit{QHeadache} SRS section from PURE~\cite{ferrari2017pure}. The section combines two distinct claims: that the product uses the Qt graphical library and that it must run on all operating systems supported by Qt. A section-level baseline treats both claims as a single generic pre-SRS artifact, obscuring which source statement supports which generated content. In contrast, \name decomposes the section into atomic statements and routes them to typed proxy artifacts: the Qt-library claim becomes a \textit{System Context and Constraints Note}, while the operating-system claim becomes a \textit{Quality Attribute Constraints Note}. This example motivates our atom-centric design, where generated artifacts remain traceable to individual source claims rather than only to coarse section-level passages.

To empirically study this gap, we use \name as a controlled, reproducible pipeline to generate synthetic pre-SRS artifacts from existing SRS documents with explicit provenance metadata. \name serves as the experimental instrument for observing how information changes as legacy SRS content is decomposed, routed, transformed into typed artifacts, and later used for reconstruction. \name uses each SRS as source evidence and operates in three stages. First, it prepares document-level SRS content for fine-grained generation by normalizing each document, splitting it into section-level chunks, and decomposing each chunk into source-grounded atomic statements. Second, it assigns artifact types to atoms using multi-LLM plurality voting followed by human-in-the-loop (HITL) review. Third, it generates typed pre-SRS artifacts using tag-specific prompts, artifact standards, and iterative quality checks. Across these stages, \name assigns stable identifiers to source sections, chunks, atoms, tags, and generated artifacts, and stores their mappings as provenance metadata. We evaluate \name at two levels: at the atom level, we study faithfulness, atomicity, downstream utility, source-to-atom coverage, factual grounding, and information loss; at the artifact level, we study faithfulness, support, contextuality, structural alignment, sparsity, and refinement needs, using judge models selected against reconciled human annotations. We also benchmark candidate LLM classifiers on PROMISE and use the best-performing models in the plurality-voting setup. We use \name as a controlled pipeline for studying how LLMs transform legacy SRS content into intermediate pre-SRS artifacts and how information changes across decomposition, routing, generation, and reconstruction. This framing allows us to evaluate not only the generated outputs, but also the practical tradeoffs that arise when public SRSs are used as seed evidence for synthetic RE resources.

We evaluate \name on seven SRS documents derived from the PURE dataset. The results show that \name produces atoms that are generally faithful and well formed: median AlignScore values are typically between 0.96 and 0.99, unsupported rates remain below 3\% for most dataset--model pairs, and final atom quality scores range from 4.34 to 4.85 under Prometheus. Generated artifacts also remain strongly grounded in their source atoms, with AlignScore values typically between 0.80--0.94 and judge pass rates near 100\% after refinement, although Prometheus reveals stricter variation in artifact quality, with pass rates ranging from 54.8\% to 97.1\%. Finally, our downstream SRS reconstruction case study shows that artifact-backed atoms remain semantically recoverable from generated SRSs, with SBERT means ranging from 0.690 to 0.753 and AlignScore medians ranging from 0.760 to 0.842. This study makes the following contributions:

\begin{itemize}
    \item We empirically characterize LLM-based atomization of legacy SRSs across seven selected PURE documents, showing that high atom-level faithfulness (median AlignScore 0.96--0.99) can coexist with wide variation in source coverage (35.6--93.5\%).

    \item We provide a stage-wise analysis of information loss across the SRS-to-atoms-to-artifacts transformation, finding that many losses reflect evidence-granularity issues rather than hallucination or parsing failure; downstream reconstruction is evaluated separately as a recoverability case study.

    \item We show that, in our setting, LLM semantic routing is more reliable than fine-grained requirement classification: Claude reaches human agreement of $\kappa=0.83$ for semantic tags and $\kappa=0.80$ for requirements, while GPT-4o/GPT-OSS agreement drops to $\kappa=0.31$ for requirement classification.

    \item We present a human-calibrated judge-selection procedure for LLM-as-a-judge evaluation of RE artifacts, based on 291 annotated triples with inter-annotator agreement of $\kappa=0.84$ for atoms and $\kappa=0.97$ for artifacts.

    \item We release a traceable replication package with processed SRSs, generated atoms and artifacts, prompts, codebooks, traceability metadata, and evaluation outputs~\cite{ReqGenXZenodo}.
\end{itemize}

The remainder of this paper is organized as follows. Section~\ref{sec:approach} describes the \name\ framework. Section~\ref{sec:experimantaldesign} presents the experimental design, and Section~\ref{sec:results} reports the main results and evaluation, including the downstream SRS reconstruction case study. Section~\ref{sec:lessons-learned} discusses empirical insights, including routing agreement, judge selection, information retention, revision behavior, and artifact-generation failures. Section~\ref{sec:related-works} reviews related work, Section~\ref{sec:threats} discusses threats to validity, and Section~\ref{sec:conclusion} concludes the paper.

\section{Study Pipeline}
\label{sec:approach}
\label{overview}
\begin{figure*} [t]
    \centering
    \includegraphics[width=1\linewidth]{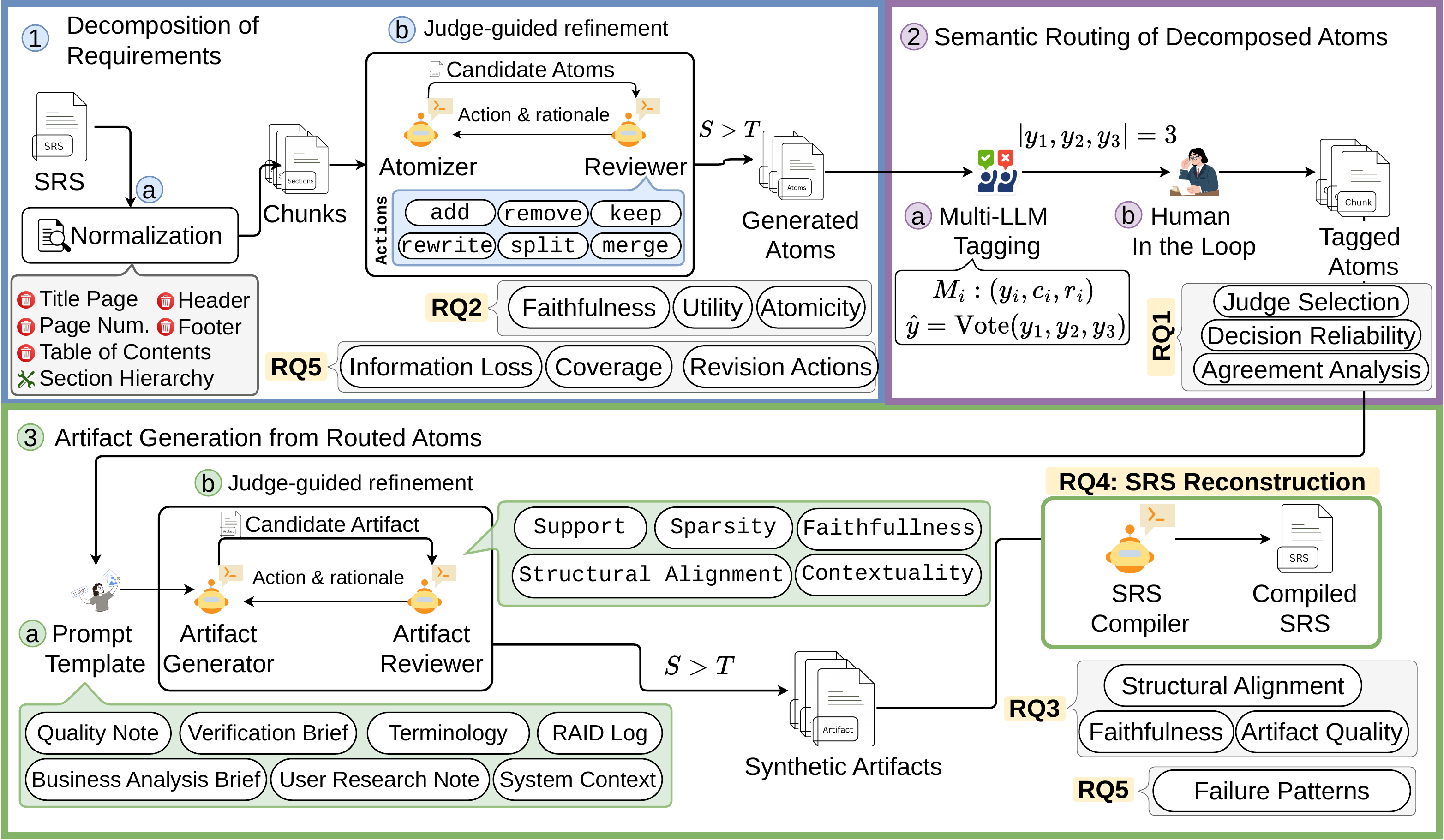}
    \caption{End-to-end \name study pipeline for deriving and evaluating traceable synthetic pre-SRS artifacts from legacy SRS documents.}
    \label{fig:overview}
\end{figure*}

\textbf{Terminology.} We use \emph{section} for a logical SRS unit after normalization, \emph{chunk} for the text associated with that section, and \emph{atom} for a minimal, source-supported proposition extracted from a chunk. An atom should express one coherent statement and remain interpretable in isolation, following the requirements-quality notion of atomicity as avoiding conflation of multiple concerns~\cite{genova2013framework}. We describe \name as the study pipeline used to operationalize our empirical analysis, rather than as a deployment-ready requirements engineering tool. Figure~\ref{fig:overview} summarizes the \name workflow for deriving standards-inspired, traceable pre-SRS artifacts from legacy SRS documents. The pipeline consists of three stages: \circleB{1} \textit{decomposition of requirements}, which produces source-grounded atoms from SRS sections; \circleB{2} \textit{semantic routing of decomposed atoms}, which maps atoms to artifact types; and \circleB{3} \textit{artifact generation and review}, which generates and refines typed pre-SRS artifacts.

\subsection{Decomposition of Requirements}

The pipeline begins with an SRS document, typically available as a PDF. Because such documents are designed for presentation rather than structured processing, we first convert each document into a text-based representation (e.g., Markdown) that preserves headings, lists, and basic layout structure. Following prior work on extracting and restructuring software documents~\cite{bacchelli2011extracting,okamoto2025restructuring}, we normalize the parsed output by removing non-requirements material such as cover pages, tables of contents, headers, footers, page numbers, broken line fragments, and other layout artifacts. Pattern-based rules repair common PDF-conversion errors, including broken section headers, missing hierarchy levels, and merged or fragmented sections. We then perform a lightweight manual validation against the original SRS to ensure that requirement content has not been removed and that parsing artifacts have not been introduced. The normalized SRS is segmented into section-wise chunks using the restored section hierarchy, and each chunk retains its section identifier, title, and source-document metadata for traceability. Each chunk is decomposed into atomic statements using an LLM-based atomizer. The atomizer is instructed to separate independent propositions while preserving necessary qualifiers, conditions, bounds, and scope information. Each atom must satisfy three criteria: \emph{faithfulness}, meaning every claim is directly supported by the source chunk; \emph{atomicity}, meaning the atom expresses one usable, self-contained proposition rather than a bundle of separable ideas or metadata-like fragment; and \emph{downstream utility}, meaning the atom is specific enough to seed a meaningful pre-SRS artifact~\cite{ronanki2023investigating,chen2024dense,hosseini2024scalable}. Because LLM-generated requirements may still be ambiguous, infeasible, or unsupported~\cite{ronanki2023investigating}, a separate judge LLM evaluates each atom set using a structured rubric. The judge scores faithfulness, atomicity, and utility, checks the set for redundancy, overlap, bad splitting or bundling, missing ideas, incoherent decomposition, and unnecessary structural atoms, and returns actions such as \texttt{keep}, \texttt{rewrite}, \texttt{split}, \texttt{merge}, \texttt{remove}, or \texttt{add}. Atom sets that fail the acceptance criterion are revised before proceeding to semantic routing. 

\subsection{Semantic Routing of Decomposed Atoms}
After decomposition, \name assigns each atom a semantic tag using an IEEE 29148-inspired schema~\cite{8559686,zhao2021natural,hey2020norbert}. Routing is needed because different SRS statements support different pre-SRS artifacts: for example, purpose and scope statements, user characteristics, definitions, assumptions, functional behavior, and quality constraints require different artifact structures. Table~\ref{tab:routingprompts} summarizes the resulting mapping from atom tags to artifact families.
\name uses an independent proposal--then--aggregate strategy for semantic routing. The atomizer first proposes an initial tag, after which three LLM agents independently label each atom without seeing peer outputs. The final tag is selected by plurality vote~\cite{wang2022self,zhu2025conformity}; unresolved cases are assigned through human-in-the-loop (HITL) review so that each atom has one canonical label before artifact generation.
\begin{table}[t]
    \centering
    \small
    \caption{Routing of atom tags to pre-SRS artifact types.}
    \label{tab:routingprompts}
    \setlength{\tabcolsep}{4pt}
    \begin{tabularx}{\linewidth}{>{\raggedright\arraybackslash}p{5cm} >{\raggedright\arraybackslash}X}
    \toprule
    \textbf{IEEE 29148 Artifact} & \textbf{Atom Tags} \\
    \midrule
    Business Analysis Brief & Purpose, Scope, Product Functions \\
    User Research & User Characteristics \\
    System Context and Constraints & Product Perspective, Limitations, Supporting Information, Other \\
    Use Case Brief & Functional \\
    Quality Attribute Constraints & Availability, Fault Tolerance, Operability, Performance, Portability, Scalability, Security, Legal, Usability, Look and Feel, Maintainability \\
    Acceptance Criteria Sheet & Verification \\
    Terminology and References & Definitions, References, Acronyms and Abbreviations \\
    RAID Log Entry & Assumptions and Dependencies \\
    \bottomrule
    \end{tabularx}
\end{table}
Atoms tagged as \texttt{requirements} undergo a second classification stage because the requirement label is too coarse for routing. \name classifies these atoms as \textit{functional}, \textit{quality}, or \textit{other}. Functional atoms are routed to \textit{Use Case Briefs}, while quality atoms are routed to \textit{Quality Attribute Constraints}. For quality atoms, \name further resolves the corresponding PROMISE-NFR subtype, such as \textit{availability}, \textit{fault tolerance}, \textit{performance}, \textit{portability}, \textit{security}, or \textit{usability}~\cite{li2024desiree,shirabad2005promise}. These subtypes guide how the quality constraint is expressed within the artifact, rather than changing the artifact family itself. The same plurality-voting and human-review process is used for this second classification stage.

\subsection{Artifact Generation and Review}
After semantic routing, \name converts each validated atom into a pre-SRS artifact using an artifact-specific template, such as a use case, quality constraint, or context note. The \texttt{atom\_text} is treated as the primary source of truth: all substantive artifact content must be supported by the atom, while metadata such as the semantic tag, section title, and chunk context is used only for formatting or minor disambiguation. Each artifact retains a traceable identifier linking it back to the source atom and its location in the original SRS. The generator returns an \texttt{artifact\_text} instance, which \name post-processes to remove formatting leakage, residual rationale text, and unintended prompt artifacts. The templates are intentionally conservative: they prioritize faithful normalization over abstraction, prohibit unsupported enrichment, omit unsupported fields, and allow empty outputs when the atom does not support a valid instance of the target artifact type. To improve artifact quality, \name applies an iterative review loop using a rubric-based judge LLM. The judge evaluates each artifact on five dimensions: \emph{faithfulness}, whether the artifact preserves the atom's meaning; \emph{support}, whether all artifact content is grounded in the atom; \emph{contextuality}, whether limited section context is used appropriately without adding new claims; \emph{structural alignment}, whether the artifact follows the expected type-specific structure; and \emph{sparsity}, whether it avoids filler, placeholders, and unsupported template sections. Artifacts that meet the acceptance threshold are retained; failed artifacts are revised using judge feedback until they pass or the refinement budget is reached. This review loop helps prevent fluent but unsupported outputs, including implicit assumptions, unnecessary structure, and template-driven filler. The eight artifact families in Table~\ref{tab:routingprompts} provide compact coverage of the main information roles underlying an SRS while remaining grounded in established requirements and architecture documentation practices~\cite{cockburn1998basic,robertson2000volere,8559686,brennan2009guide,starke2019arc42,392555}.

\subsection{Traceability and Provenance Model}
\name maintains provenance metadata across the generated artifacts so that each output can be related back to its source evidence. During chunking, each section-wise chunk stores document-level and location metadata, including the document identifier, source file, section label, section identifier, and section title. During artifact generation, \name combines the document identifier with the atom identifier to create a stable artifact-level trace identifier. This identifier is stored with the generated artifact together with the source atom, semantic tag, requirement-type labels, section metadata, and supporting chunk context. As a result, the pipeline records a practical trace chain from the SRS document and section to chunk, atom, semantic tag, generated artifact, and, in the reconstruction case study, matched SRS sentence windows. We use these links to support grounding checks, information retention analysis, and downstream recoverability evaluation. We do not claim that this metadata provides a complete industrial traceability solution; rather, it provides the provenance needed to audit LLM-generated artifacts against their source SRS evidence.

\section{Experimental Design}
\label{sec:experimantaldesign}

We use \name as a controlled pipeline to study how LLMs perform across the transformations required to derive traceable synthetic pre-SRS artifacts from legacy SRS documents. The study covers five dimensions: intermediate RE decisions, atomic decomposition, artifact generation, downstream reconstruction, and sources of information loss.

\begin{itemize}
    \item \textbf{RQ1:} To what extent can LLMs support RE-specific intermediate decisions, including semantic classification and quality judgment?
    \item \textbf{RQ2:} Can LLMs decompose legacy SRS sections into faithful, atomic, and downstream-usable statements while preserving source information?
    \item \textbf{RQ3:} To what extent can LLMs generate grounded, sparse, and structurally valid pre-SRS artifacts from routed atoms?
    \item \textbf{RQ4:} Can LLM-generated pre-SRS artifacts support downstream SRS reconstruction?
    \item \textbf{RQ5:} Where is information lost across the transformation process, and which revision or failure patterns explain these losses?
\end{itemize}

\subsection{Dataset}
\label{dataset}

We construct our benchmark from the PURE dataset~\cite{ferrari2017pure} using the document-selection protocol of Okamoto et al.~\cite{okamoto2025restructuring}. Because \name operates on text sequences, we select SRS documents whose requirements content can be represented as ordered textual sections, paragraphs, and lists. We retain only documents explicitly identified as Software Requirements Specifications (SRSs), excluding other specification types such as functional or system requirements specifications. We also exclude SRSs with substantive figures or non-revision-history tables, since these elements do not provide a stable linear reading order for sentence segmentation, atomization, and downstream faithfulness evaluation. Revision-history tables are retained because they can be normalized into linear text without changing their meaning. After filtering, we retain seven PURE SRS documents: Get Real, Library, Puget Sound, QHeadache, VUB, Libra, and Home. In total, the benchmark contains 31,454 words, 1,408 sentences, and 369 sections. This filtering favors SRSs whose requirements content can be represented as linear text; therefore, our results may understate failures that would arise for figure-heavy, table-heavy, or multimodal SRSs.

\subsubsection{Implementation and Large Language Models}
Our pipeline was implemented in Python 3.11, and all LLM interactions were orchestrated via LangChain. We ran all experiments on a system with a 28-core Intel\textsuperscript{\textregistered} Xeon\textsuperscript{\textregistered} W-3465X CPU (2.50\,GHz), 256\,GB RAM, and an NVIDIA RTX 6000 Ada Generation GPU, running Ubuntu~24.04.2~LTS. We set the temperature to 1.0 to encourage output diversity, aiming to generate varied artifacts rather than deterministic rewrites. In this work, we use LLMs for four main tasks: atom generation, semantic tagging, artifact generation, and review. Our experiments include both closed-source and open-source models from multiple families and parameter scales, including Claude Sonnet 4.6, GPT-4o, GPT-OSS (20.9B), Gemma 3 (27.4B), Phi-4 (14.7B), Mistral (7.2B), Llama 3.1 (8.0B), Qwen3 (30.5B and 8.2B), and LLaMA 3 (8.0B). This model set enables us to compare behavior across both generation and evaluation roles.

\subsection{Evaluation Metrics}
\label{evaluation}
\textbf{AlignScore:} We use AlignScore~\cite{zha2023alignscore} to measure whether generated content is semantically supported by its source. For atom evaluation, the source chunk serves as the context, and the generated atom as the claim. For artifact evaluation, the source atom serves as the context, and the generated artifact as the claim. We use 0.80 as the support threshold when reporting unsupported rates.

\textbf{Judge LLM and Prometheus Evaluation:} We use a rubric-based judge LLM to evaluate atom and artifact quality on a 1--5 scale, treating scores $\geq 4$ as passing. Atoms are judged for faithfulness, atomicity, and downstream utility; artifacts are judged for faithfulness, unsupported content, contextual use, structural alignment, and sparsity. We also use Prometheus~\cite{kim2024prometheus} as an external rubric-based evaluator to provide a stricter complementary assessment of atom and artifact quality.

\textbf{Embedding-Based Information Retention}: We measure source information retention using cosine similarity over \texttt{text-embedding-3-large}. Source SRS text is split into overlapping two-sentence units, and each unit is matched to the final atoms. A source unit is treated as covered if its best-match similarity is at least 0.75. We use this setup to estimate source coverage after atomization and stage-wise retention across SRS$\rightarrow$Atoms and Atoms$\rightarrow$Artifacts transformations.

\textbf{Downstream SRS Reconstruction Metrics:} For the reconstruction case study, we evaluate whether accepted artifact-backed atoms are recoverable from the generated SRS. We segment the generated SRS into overlapping sentence windows and match each source atom to its closest generated SRS window using Sentence-BERT (SBERT) cosine similarity~\cite{reimers2019sentence}. We then apply AlignScore with the generated SRS window as context and the source atom as the claim. SBERT measures semantic similarity, while AlignScore estimates factual support.

\section{Results and Evaluation}
\label{sec:results}

\subsection{RQ1: Reliability of LLM-Based RE Decisions}
\label{subsec:rq1-decision-reliability}
\begin{table}[t]
\centering
\small
\caption{Performance of candidate LLMs on the PROMISE dataset for plurality voting selection. Metrics are reported as support-weighted per-class precision, recall, and F1; weighted F1 is averaged from per-class F1 values.}
\label{tab:routing-model-selection}
\begin{tabular}{lccc}
\toprule
\textbf{Model} & \textbf{Precision} & \textbf{Recall} & \textbf{F1} \\
\midrule
Claude Sonnet 4.6 & 0.80 & \underline{0.84} & \textbf{0.81} \\
GPT-4o            & 0.78 & \textbf{0.86} & \textbf{0.81} \\
\midrule
GPT-OSS           & \underline{0.84} & 0.68 & \underline{0.70} \\
Phi-4             & \underline{0.84} & 0.55 & 0.59 \\
Qwen3 (30B)       & \textbf{0.92} & 0.43 & 0.50 \\
Qwen3 (8B)        & 0.80 & 0.46 & 0.50 \\
LLaMA 3 (8B)      & 0.44 & 0.37 & 0.36 \\
\bottomrule
\end{tabular}
\begin{flushleft}
\footnotesize
\textbf{Note:} Bold values indicate the best value in each metric column; underlined values indicate the second-best value. Ties are marked for all tied values.
\end{flushleft}
\end{table}
\begin{figure}[t]
    \centering
    \includegraphics[width=1\textwidth]{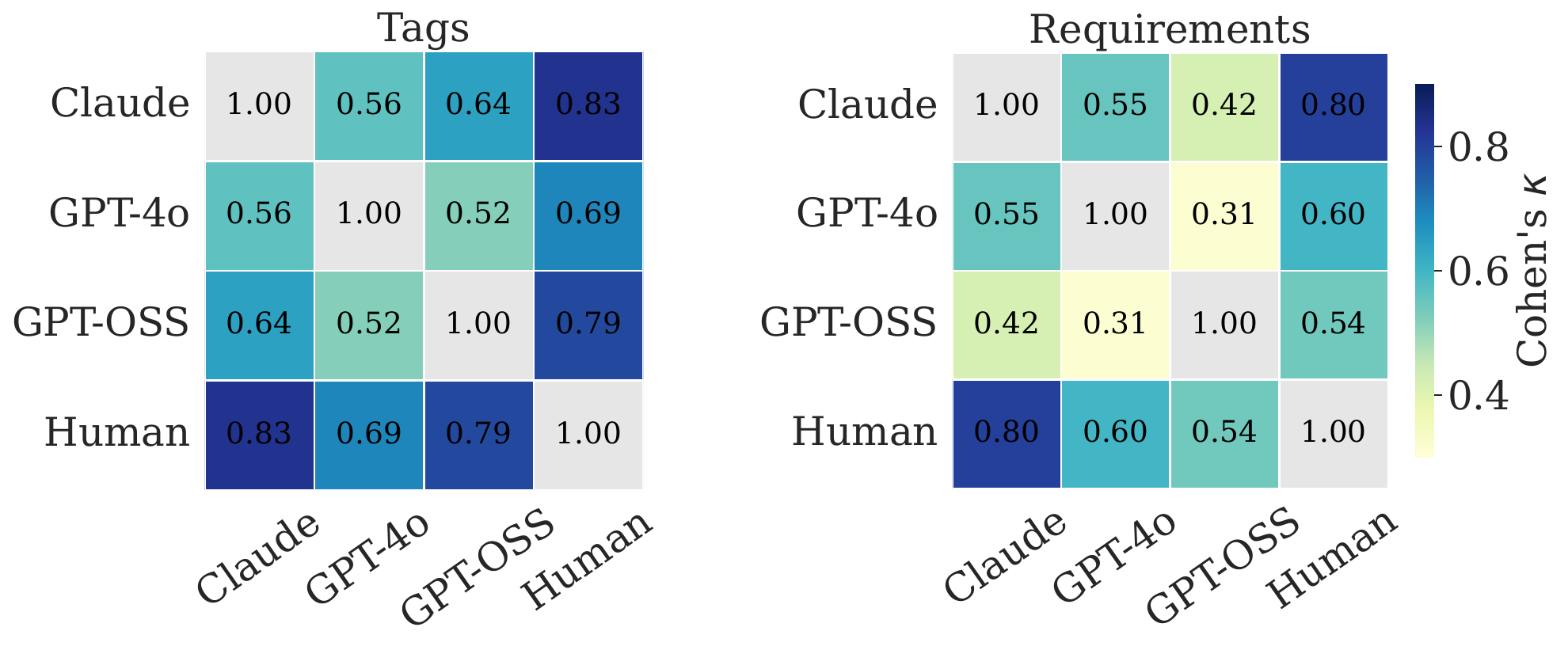}
    \caption{Pairwise Cohen's $\kappa$ agreement among LLM-generated and human-reviewed labels for semantic tag assignment and requirement classification. Diagonal cells denote self-agreement, while off-diagonal cells show agreement between annotator/model pairs.}
    \label{fig:llm-kappa}
\end{figure}
\begin{table}[t]
\centering
\caption{Comparison of candidate judge LLMs for atom- and artifact-level assessment against reconciled human labels. F1 is higher-is-better. FPR and FNR are lower-is-better.}

\label{tab:judge-comparison}
\begin{tabularx}{\columnwidth}{
l
>{\centering\arraybackslash}X
>{\centering\arraybackslash}X
>{\centering\arraybackslash}X
>{\centering\arraybackslash}X
>{\centering\arraybackslash}X
>{\centering\arraybackslash}X
}
\toprule
& \multicolumn{3}{c}{\textbf{Atoms}} 
& \multicolumn{3}{c}{\textbf{Artifacts}} \\
\cmidrule(lr){2-4}
\cmidrule(lr){5-7}
\textbf{LLM} & \textbf{F1} $\uparrow$ & \textbf{FPR} $\downarrow$ & \textbf{FNR} $\downarrow$
& \textbf{F1} $\uparrow$ & \textbf{FPR} $\downarrow$ & \textbf{FNR} $\downarrow$ \\
\midrule
Gemma 3   & \underline{0.80} & 0.33 & 0.20 & \textbf{0.88} & \textbf{0.29} & 0.13 \\

GPT-4o    & 0.77 & \underline{0.28} & 0.27 & \textbf{0.88} & 0.63 & \underline{0.03} \\

Phi-4     & 0.79 & 0.46 & \underline{0.16} & \underline{0.87} & \underline{0.47} & 0.09 \\

Mistral   & 0.74 & 0.89 & \textbf{0.12} & 0.84 & 1.00 & \textbf{0.01} \\

Llama 3.1 & 0.66 & \textbf{0.23} & 0.44 & 0.83 & 0.90 & 0.04 \\

GPT-OSS   & \textbf{0.81} & 0.49 & \textbf{0.12} & 0.76 & 0.77 & 0.21 \\
\bottomrule
\end{tabularx}
\begin{flushleft}
\footnotesize
\textbf{Note:} Bold values indicate the best value in each metric column; underlined values indicate the second-best value. Ties are marked for all tied values.
\end{flushleft}
\end{table}

RQ1 examines whether LLMs can support two intermediate decisions required for traceable artifact generation: semantic routing and quality judgment. Semantic routing determines which artifact family an atom should support, while quality judgment determines whether generated atoms and artifacts are acceptable for downstream use. We evaluate these decisions through routing-model screening, agreement with human-reviewed labels, and human-grounded judge selection. For routing, we first screen candidate LLMs on the re-labeled PROMISE-NFR dataset\footnote{\url{https://zenodo.org/records/14330485}} as a proxy benchmark for requirement-style classification. Because PROMISE-NFR does not cover the full \name routing schema, Table~\ref{tab:routing-model-selection} should be interpreted as evidence of general requirement-classification ability rather than direct routing accuracy. Claude Sonnet 4.6 and GPT-4o achieve the strongest weighted F1 among closed-source models, while GPT-OSS is the strongest open-source model; we therefore use Claude Sonnet 4.6, GPT-4o, and GPT-OSS 20B as complementary voters for plurality-based routing. We then evaluate routing on the actual \name schema using human-reviewed labels. The agreement analysis includes 2,054 semantic-tag instances and 1,177 requirement-label instances; human review corrected 353 semantic tags (17.2\%) and 58 requirement labels (4.9\%). Figure~\ref{fig:llm-kappa} shows that agreement is consistently higher for semantic tags than for requirement labels, indicating that high-level routing is more stable than fine-grained requirement classification. Claude has the strongest agreement with human-reviewed labels, reaching $\kappa=0.83$ for tags and $\kappa=0.80$ for requirements, while model-to-model agreement is weakest for requirement classification, where GPT-4o and GPT-OSS fall to $\kappa=0.31$. For quality judgment, we compare candidate judge LLMs against reconciled human annotations from 291 triples sampled from 1,195 Claude Sonnet 4.6-generated artifact instances. Two annotators scored atom and artifact quality using a shared 1--5 rubric, achieving Cohen's $\kappa=0.84$ for atoms and $\kappa=0.97$ for artifacts; disagreements were adjudicated to create gold-standard labels. Table~\ref{tab:judge-comparison} compares candidate judges using F1, false positive rate (FPR), and false negative rate (FNR). Since false acceptance can propagate low-quality atoms or artifacts into later stages, we prioritize low FPR. Gemma 3 provides the best overall tradeoff, with strong F1 for atoms (0.80) and artifacts (0.88) while keeping FPR lower than alternatives such as GPT-OSS for atoms and GPT-4o for artifacts. We therefore use Gemma 3 as the automated judge for large-scale evaluation.

\finding{LLMs can support RE-specific intermediate decisions, but reliability varies by decision type. Semantic tags are more stable than fine-grained requirement labels: Claude reaches $\kappa=0.83$ for tags and $\kappa=0.80$ for requirements, while model-to-model agreement drops to $\kappa=0.31$ for GPT-4o vs. GPT-OSS on requirement classification. For quality judgment, Gemma 3 provides the best overall tradeoff, with F1 scores of 0.80 for atoms and 0.88 for artifacts while maintaining lower false-positive rates than the strongest alternatives.}

\subsection{RQ2: Quality of Generated Atoms}
\begin{figure}[t]
    \centering
    \includegraphics[width=1\textwidth]{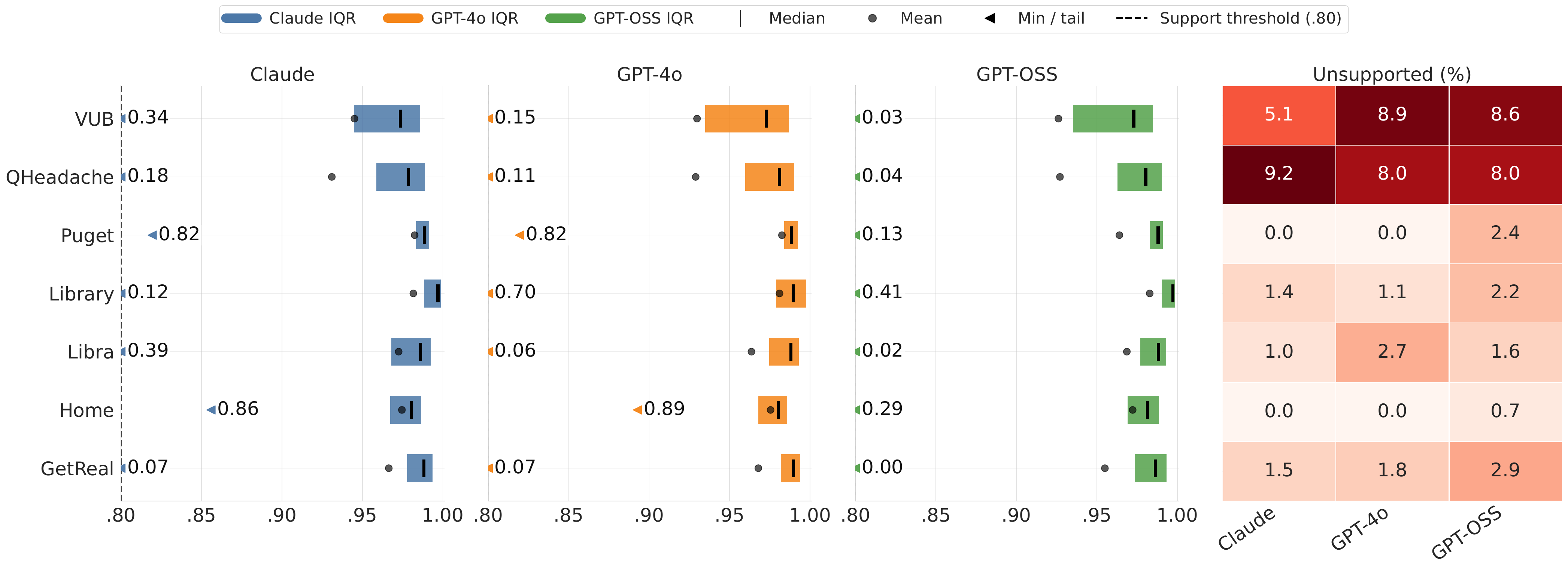}
    \caption{
    Chunk-level atom faithfulness across datasets and atomizer models. Most atomized outputs achieve high central AlignScore values, with medians generally above 0.95, but several dataset--model pairs exhibit low-scoring tail cases. Unsupported rates are computed using the AlignScore support threshold of 0.80 and show that high typical faithfulness can still coexist with occasional unsupported atoms.
    }
    \label{fig:alignScore}
\end{figure}
\begin{figure}[t]
    \centering
    \includegraphics[width=1\textwidth]{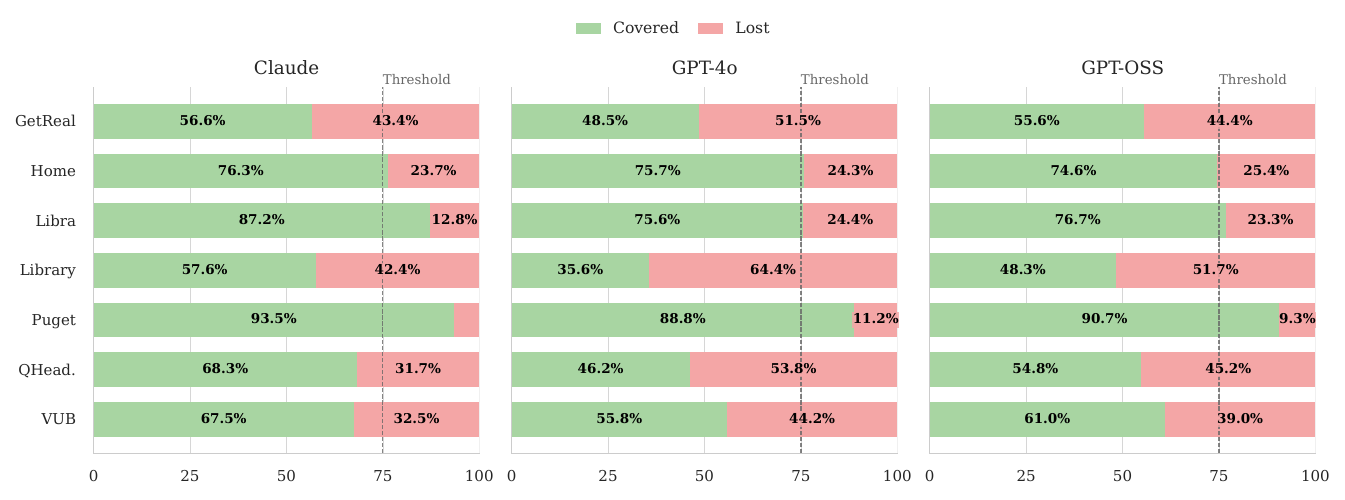}
    \caption{Information coverage and loss across datasets using overlapping window-2 source units with a similarity threshold of 0.75. Each horizontal stacked bar shows the percentage of covered and lost source information units for a given model and dataset. }
    \label{fig:coverage}
\end{figure}
\begin{table}[t]
\centering
\caption{
Final atom quality across SRSs and atomizer models. Atomicity and usability pass rates report the percentage of final atoms receiving judge scores $\geq 4$; average Judge (Gemma 3) and Prometheus scores are reported on a 1--5 scale.
}
\label{tab:atom-quality}
\resizebox{\textwidth}{!}{%
\begin{tabular}{@{}lrrr|rrr|rrr|rrr@{}}
\toprule
\multirow{2}{*}{SRS} &
  \multicolumn{3}{c|}{Atomicity Pass \%} &
  \multicolumn{3}{c|}{Usability Pass \%} &
  \multicolumn{3}{c|}{Avg. Judge Score} &
  \multicolumn{3}{c}{Avg. Prometheus Score} \\ \cmidrule(l){2-13} 
 &
  Claude &
  GPT-4o &
  GPT-OSS &
  Claude &
  GPT-4o &
  GPT-OSS &
  Claude &
  GPT-4o &
  GPT-OSS &
  Claude &
  GPT-4o &
  GPT-OSS \\ 
  \midrule
VUB     & \textbf{98.1} & 92.5 & \underline{96.4} & \textbf{96.2} & 85.6 & \underline{89.8} & \textbf{4.84} & 4.70 & \underline{4.75} & \textbf{4.72} & \underline{4.69} & 4.67 \\
QHeadache  & \textbf{98.0} & 76.0 & \underline{87.5} & \textbf{93.9} & 70.7 & \underline{75.0} & \textbf{4.84} & 4.65 & \underline{4.66} & \underline{4.77} & \underline{4.77} & \textbf{4.80} \\
Puget Sound   & \underline{96.2} & \textbf{96.3} & 94.5 & 82.6 & \underline{84.6} & \textbf{87.4} & 4.67 & \textbf{4.69} & \textbf{4.69} & 4.34 & \textbf{4.44} & \underline{4.37} \\
Library & \textbf{90.0} & 85.6 & \underline{87.6} & \textbf{87.9} & \underline{82.2} & 80.3 & \textbf{4.87} & \underline{4.76} & 4.75 & \underline{4.81} & 4.77 & \textbf{4.85} \\
Libra   & 93.0 & \underline{94.1} & \textbf{95.3} & \underline{91.0} & 89.8 & \textbf{92.2} & 4.78 & \textbf{4.80} & \textbf{4.80} & 4.72 & \textbf{4.75} & 4.72 \\
Home    & 98.5 & \textbf{100} & \underline{99.3} & \textbf{97.0} & 90.3 & \underline{94.6} & \textbf{4.90} & \underline{4.86} & 4.85 & \underline{4.70} & 4.69 & \textbf{4.72} \\
Get Real & 95.4 & 93.0 & \textbf{95.7} & 81.5 & \underline{82.5} & \textbf{85.7} & \underline{4.64} & 4.62 & \textbf{4.73} & 4.46 & \textbf{4.60} & \underline{4.59} \\
\bottomrule
\end{tabular}%
}
\footnotesize
\flushleft
\textit{Note.} \textbf{Bold} values indicate the best-performing model for each dataset and metric group; \underline{underlined} values indicate the second-best model. Ties are marked for all tied models.
\end{table}

RQ2 evaluates whether LLMs can decompose legacy SRS sections into atoms that are faithful, atomic, and useful for downstream artifact generation. Overall, the results show that LLM-generated atoms are usually well grounded, but atomization is not equivalent to lossless extraction: usable decomposition requires balancing faithfulness, source coverage, atomicity, and downstream utility. Figure~\ref{fig:alignScore} shows that generated atoms are generally faithful to their source chunks. Across datasets and models, median AlignScore values typically range from 0.96 to 0.99, while unsupported rates remain below 3\% for most dataset--model pairs. The strongest cases occur in Puget Sound and Home, where unsupported rates are 0.0\% for Claude Sonnet 4.6 and GPT-4o, and remain low for GPT-OSS. The highest unsupported rates appear for QHeadache with Claude-generated atoms (9.2\%) and for VUB with GPT-4o and GPT-OSS atoms (8.9\% and 8.6\%, respectively). These low-scoring tails show that median faithfulness alone is insufficient: occasional unsupported atoms can still appear and may propagate into routing and artifact generation. Model-level differences are small and should be interpreted cautiously. A paired bootstrap over the seven SRSs shows that Claude and GPT-4o are not clearly separated on mean AlignScore, while Claude is only slightly higher than GPT-OSS; for unsupported rate, Claude is lower than GPT-OSS by 1.13 percentage points, but not clearly separated from GPT-4o. We therefore treat model rankings as descriptive rather than conclusive.

\finding{LLMs can decompose legacy SRS sections into faithful, atomic, and usable statements. Across datasets and models, median AlignScore values typically fall between 0.96 and 0.99, most unsupported rates remain below 3\%, and most atomicity pass rates exceed 90\%. Source-coverage results further show that decomposition quality should be interpreted together with information-retention analysis.}

Figure~\ref{fig:coverage} shows a sharper tradeoff for source coverage. Puget Sound is consistently best preserved, with 88.8--93.5\% coverage, followed by Libra with 75.6--87.2\%. In contrast, GPT-4o covers only 35.6\% of Library and 46.2\% of QHeadache. This suggests that atomization is often faithful in what it retains, but may omit or transform source information depending on the dataset and model. Some loss is expected: judge-guided revision can remove redundant, structural, vague, or low-utility content to produce cleaner atoms for downstream use. Thus, coverage should be interpreted alongside faithfulness rather than as a standalone failure signal. Table~\ref{tab:atom-quality} shows that final atoms are generally well formed after revision. Atomicity pass rates range from 76.0\% to 100.0\%, with most dataset--model pairs above 90\%, while usability pass rates range from 70.7\% to 97.0\%. Average Gemma 3 judge scores remain high, ranging from 4.62 to 4.90, and Prometheus scores range from 4.34 to 4.85. These results indicate that LLMs can produce strong atomic decompositions, but downstream usability is harder to achieve than atomicity and benefits from explicit judge-guided refinement.

\subsection{RQ3: Quality of the Generated Artifacts}
\begin{figure}[t]
    \centering
    \includegraphics[width=1\textwidth]{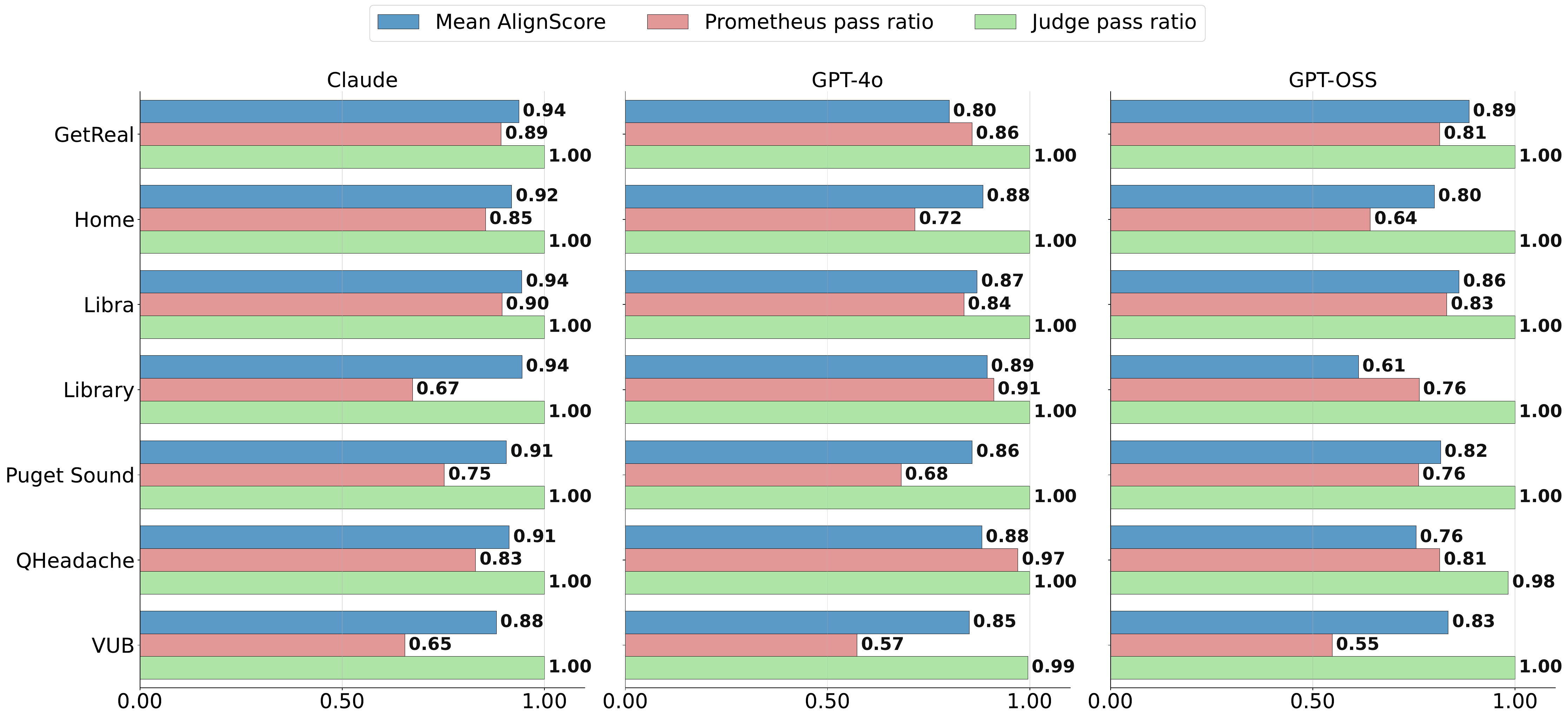}
    \caption{Artifact quality over generated artifacts. Mean AlignScore, Prometheus pass ratio, and judge pass ratio are reported on a 0--1 scale. All models show high atom-level grounding and high judge acceptance, while Prometheus reveals stricter variation in artifact completeness and quality.}
    \label{fig:artifact-quality}
\end{figure}

Our goal is to assess whether atom-derived artifacts are both grounded in their source atoms and suitable for downstream requirements generation. Since artifacts are generated through type-specific templates and refined through a judge-guided loop, we evaluate them along three complementary dimensions: atom-level grounding using mean AlignScore, acceptance under the internal judge, and stricter external quality assessment using Prometheus. All three metrics are reported on a 0--1 scale. Figure~\ref{fig:artifact-quality} evaluates whether generated artifacts remain grounded in their source atoms and whether they satisfy downstream quality requirements after refinement. Overall, mean AlignScore values remain consistently high across datasets and models, typically ranging between 0.80 and 0.94, indicating that generated artifacts are largely supported by their source atoms. Claude Sonnet 4.6 shows the strongest and most stable grounding performance, reaching 0.94 on Library, while GPT-OSS exhibits larger variation, dropping to 0.61 on the same dataset. Judge pass ratios are nearly saturated after iterative refinement, remaining close to 1.00 across almost all dataset--model pairs. This suggests that the review-and-repair loop reliably produces artifacts that satisfy the pipeline's acceptance criteria. However, Prometheus reveals substantially greater variation, with pass ratios ranging from 0.55 to 0.97. The strongest Prometheus result appears on QHeadache using GPT-4o (0.97), while lower scores occur for VUB and Home, particularly for GPT-OSS and GPT-4o. This gap indicates that atom-level grounding alone does not guarantee artifact completeness, sparsity, or structural quality. Taken together, these results show that \name\ can generate artifacts that remain strongly grounded in their source atoms while achieving high refinement success rates. At the same time, stricter rubric-based evaluation reveals that downstream artifact quality remains sensitive to dataset characteristics and model behavior, motivating multi-dimensional evaluation beyond grounding alone.

\finding{LLMs can generate grounded pre-SRS artifacts from routed atoms, with mean AlignScore values typically between 0.80 and 0.94 and judge pass ratios near 1.00 after refinement. However, stricter Prometheus pass ratios range from 0.55 to 0.97, showing that grounded artifacts still require artifact-specific quality checks for sparsity, completeness, and structural alignment.}

\subsection{RQ4: Downstream SRS Reconstruction}
\label{sec:rq4-reconstruction}
\begin{table}[t]
\centering
\caption{Downstream reconstruction retention scores. SBERT measures semantic similarity between source atoms and generated SRS, while AlignScore estimates factual support.}
\label{tab:reconstruction-retention}
\begin{tabular}{@{}lrrrrrr@{}}
\toprule
\multirow{2}{*}{Dataset} & \multicolumn{3}{c}{SBERT}     & \multicolumn{3}{c}{AlignScore} \\ \cmidrule(l){2-7} 
                         & Mean  & Median & IQR          & Mean   & Median & IQR          \\ 
\midrule
PugetSound               & 0.697 & 0.717  & [0.622--0.782] & 0.596  & 0.785  & [0.081--0.988] \\
Library                  & 0.698 & 0.720  & [0.640--0.764] & 0.643  & 0.842  & [0.307--0.962] \\
Libra                    & 0.690 & 0.694  & [0.634--0.755] & 0.551  & 0.785  & [0.040--0.974] \\
Home                     & 0.753 & 0.770  & [0.706--0.821] & 0.633  & 0.760  & [0.261--0.984] \\ 
\bottomrule
\end{tabular}%
\end{table}
RQ4 evaluates whether traceable pre-SRS artifacts can support downstream SRS reconstruction while preserving source-grounded information. We conduct this as a limited four-dataset case study because complete SRS compilation requires an additional generation, evaluation, and traceability-matching pipeline. The goal is not practitioner-ready SRS generation, but to test whether source atoms remain factually recoverable after artifact-based reconstruction. We use Claude Sonnet 4.6-generated artifacts as input because Claude showed strong upstream performance, and GPT-4o as a separate compiler model to avoid using the same model for artifact generation and reconstruction. GPT-4o compiles the artifacts into SRS drafts under ISO/IEC/IEEE 29148-inspired constraints while preserving artifact-level traceability. The generated requirements are then refined through a quality-assurance loop for ambiguity, singularity, verifiability, and syntactic correctness. To evaluate retention, we trace each generated SRS back to its artifact-backed source atoms. We segment the generated SRS into overlapping sentence windows, match each source atom to its closest window using SBERT cosine similarity, and apply AlignScore with the matched window as context and the source atom as the claim.

\finding{LLM-generated pre-SRS artifacts can support partial downstream SRS reconstruction, but semantic similarity overestimates factual recoverability. The reconstructed SRSs should be interpreted as traceable drafts for evaluation, not as complete or practitioner-ready specifications.}

Table~\ref{tab:reconstruction-retention} reports retention scores for SRSs compiled by GPT-4o from Claude Sonnet 4.6-generated artifacts. Across 473 accepted artifact-backed atoms, SBERT means range from 0.690 to 0.753, indicating moderate-to-high semantic recoverability. AlignScore medians remain high, ranging from 0.760 to 0.842, but lower means and wider IQRs show that factual support is more variable than semantic similarity. These low-support cases may reflect abstraction, rephrasing, or additional content introduced during SRS compilation rather than direct omission alone.

\subsection{RQ5: Information Loss and Failure Patterns}
\label{sec:rq5-information-loss}
\begin{figure}[t]
    \centering
    \includegraphics[width=1\textwidth]{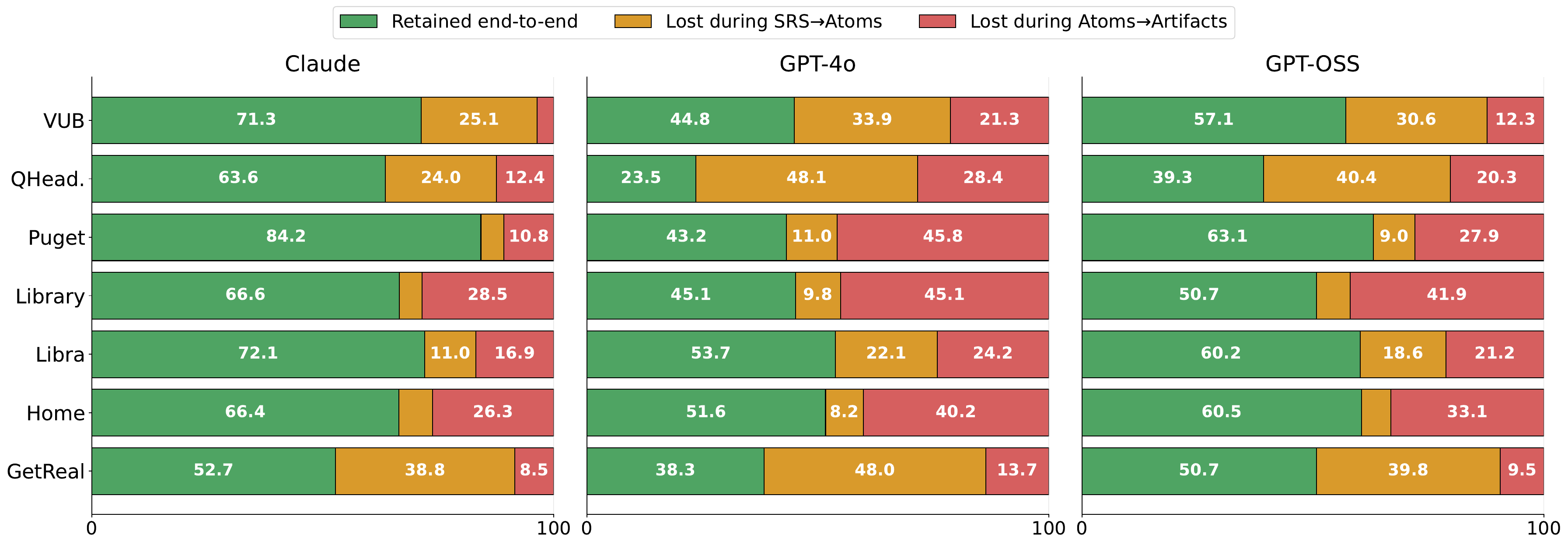}
    \caption{Stage-wise information retention across the \name pipeline. Green indicates information retained end-to-end, while yellow and red indicate estimated information loss during atomization and artifact generation, respectively.}
    \label{fig:information-retention}
\end{figure}

\begin{figure}[t]
    \centering
    \includegraphics[width=1\textwidth]{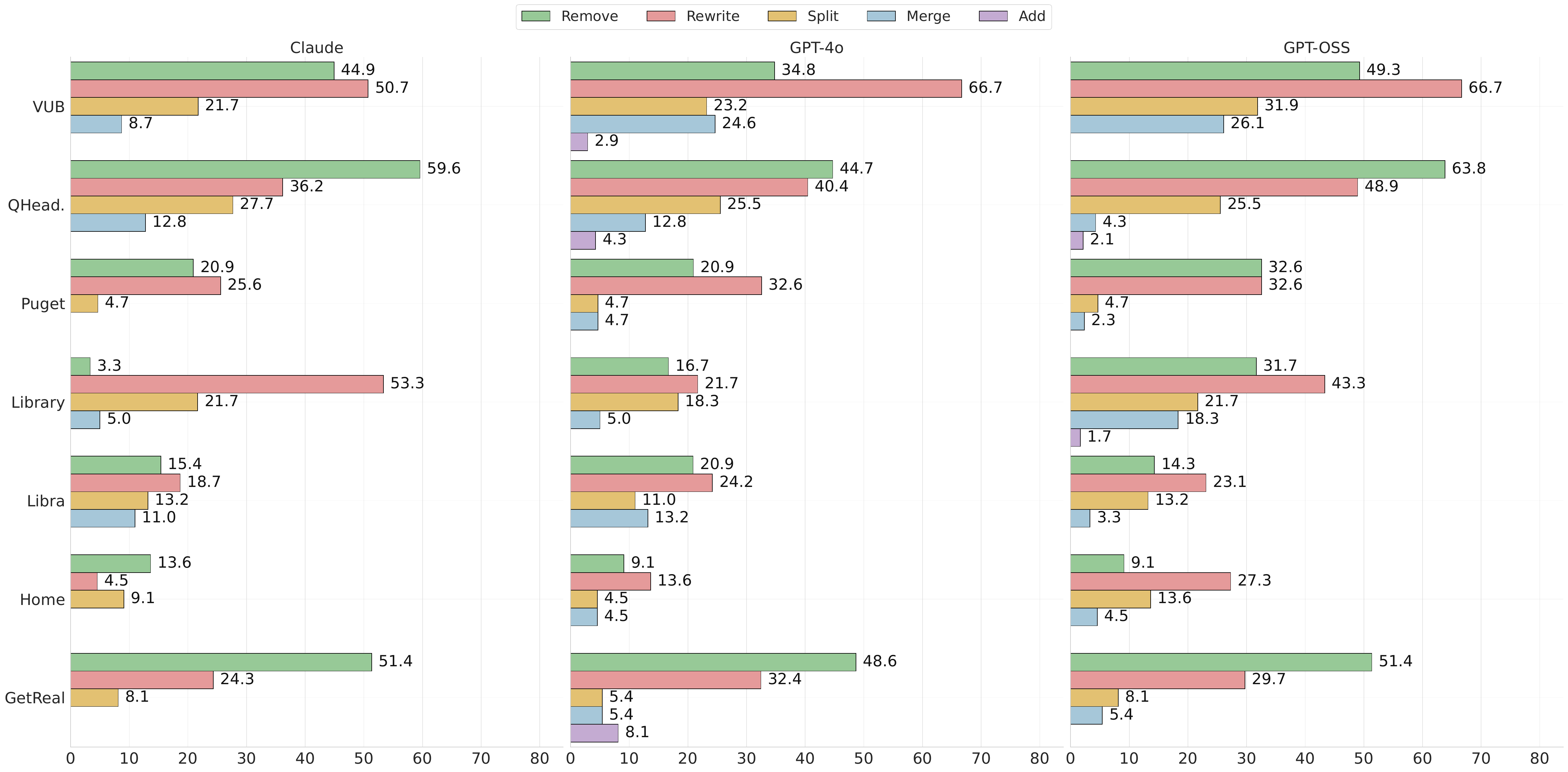}
    \caption{Revision action rates across datasets and atomizer models. Rewrite and remove actions dominate across most dataset--model pairs, while split actions remain visible for structurally dense SRS sections.}
    \label{fig:revisionRates}
\end{figure}

RQ5 examines where source information is lost across the transformation process and what revision patterns explain these losses. We analyze stage-wise retention across SRS$\rightarrow$Atoms and Atoms$\rightarrow$Artifacts, and summarize descriptive diagnostics from judge-revision logs and artifact-generation traces. Information loss is not concentrated in a single step. Claude retains the most source information end-to-end, preserving 52.7--84.2\% of source content, while GPT-4o shows the largest cumulative loss, dropping to 23.5\% retention on QHeadache; GPT-OSS falls between these ranges, retaining 39.3--63.1\%. Loss also varies by dataset: QHeadache and Get Real lose substantial information during SRS$\rightarrow$Atoms conversion, suggesting that implicit or weakly structured statements are difficult to stabilize as atoms, whereas Library and Puget Sound lose more information during Atoms$\rightarrow$Artifacts conversion, indicating that valid atoms may still be too narrow to instantiate constrained artifact templates. Revision and artifact diagnostics explain these patterns. Most atom-level revisions involve granularity, context, or downstream usefulness rather than unsupported hallucination. The most frequent buckets are bundled or non-atomic claims (30.4\%), context-dependent wording (20.4\%), and weak or non-actionable wording (8.5\%)~\cite{ReqGenXZenodo}. Rewrite and remove actions dominate, while split actions remain visible for dense SRS sections; merge and add actions are rare. This suggests that the atomizer more often over-produces weak or underspecified atoms than under-produces missing ones. Artifact-level failures show a related evidence-granularity problem: missing or incomplete artifacts are rarely due to parsing or generation crashes, but usually occur because the selected atom lacks enough evidence to fill the target artifact structure. For Claude, sparse artifacts are often associated with \textit{chunk-only information}, while for GPT-4o and GPT-OSS the dominant issue is \textit{template fields omitted}.

\finding{Information loss is stage-specific and often reflects evidence granularity rather than generation failure. Decomposition can discard implicit, redundant, or weakly actionable source content, while artifact generation can fail when a valid atom is too narrow to instantiate a complete typed artifact.}

\section{Lessons Learned }
\label{sec:lessons-learned}

\circleB{1} \textit{LLM-based RE decisions need uncertainty-aware evaluation.}
RQ1 shows that LLMs can support semantic routing and quality judgment, but reliability varies across decision types. This matters because routing determines which artifact an atom becomes, while judging determines which outputs are accepted for later stages. Disagreement should not be treated only as noise; it can identify ambiguous requirements, weak artifact boundaries, and cases where human review is most valuable. Large Language Models for Requirements Engineering (LLM4RE) researchers should therefore evaluate routing confidence, disagreement patterns, and judge false-acceptance risk instead of reporting only aggregate accuracy. Practitioners should review low-confidence routing decisions and judge-approved outputs rather than treat them as reliable artifacts.

\circleB{2} \textit{Not every SRS statement is immediately artifact-ready.}
RQ2 shows that LLM-generated atoms can be faithful and well formed, but source coverage varies across datasets and models. This reflects a property of legacy SRSs: they mix concrete requirements with background context, vague notes, structural fragments, and partially specified ideas. Some statements can directly seed pre-SRS artifacts, while others need added context or aggregation with neighboring atoms. LLM4RE researchers should therefore study artifact readiness: which statements should become artifacts, which should remain supporting context, and how coverage can be improved without reducing downstream utility. Practitioners should avoid forcing every SRS sentence into an artifact template; weak content may be better treated as context, clarification targets, or evidence for grouped artifacts.

\circleB{3} \textit{Artifact generation requires balancing completeness and evidence control.}
RQ3 shows that artifacts can remain grounded in their source atoms, but grounding alone does not make them complete or useful. The key tradeoff is that adding synthetic details can improve completeness but risks unsupported content, while strict atom-grounding preserves factuality but may produce sparse artifacts. Evidence-aware generation should therefore decide when to produce full artifacts, partial artifacts, or request more context. Missing fields should be treated as diagnostic signals that the source evidence is incomplete, too narrow, or not ready for artifact-level synthesis.

\circleB{4} \textit{SRS reconstruction is feasible, but factuality must be checked across artifact chains.}
RQ4 shows that LLM-generated pre-SRS artifacts can be compiled into SRS drafts with partial source recoverability. However, synthetic details introduced during artifact generation may bleed into the reconstruction, creating fluent, structurally valid SRS text that is only partially supported by the original atoms. Future work should therefore evaluate factuality across the full chain from source SRS to atoms, artifacts, and reconstructed SRS, rather than only the final document. Generated SRS drafts should be reviewed through trace links, especially when requirements rely on synthetic artifact content.

\circleB{5} \textit{The main bottleneck is evidence granularity.}
RQ5 shows that information loss is not mainly caused by parsing crashes or random generation failure; it often occurs because the available evidence is either too implicit, too bundled, or too narrow for the next transformation step. At the SRS$\rightarrow$Atom stage, weakly structured or context-dependent statements are difficult to turn into standalone atoms. At the Atom$\rightarrow$Artifact stage, valid atoms may still be too small to fill a typed artifact template without adding unsupported content. The broader lesson is that future LLM4RE systems need mechanisms for carrying and using context at the right granularity, such as grouping related atoms, preserving context-only evidence, and allowing partially specified artifacts instead of forcing every atom into a complete template. In practice, analysts should treat missing or sparse artifacts as signals that the source evidence needs clarification, grouping, or richer contextual support before it can be used for requirements generation.

\section{Related Work}
\label{sec:related-works}

Recent work shows that LLMs can support requirements engineering (RE) activities such as elicitation, user-story generation, acceptance-criteria generation, functional design specification, and SRS drafting~\cite{ronanki2023investigating,quattrocchi2026can,spijkman2025llm,pasquale2025exploring,lazik2025good,krishna2024using,xie2025effective,zhu2025reqinone}. Broader Large Language Models for Software Engineering (LLM4SE) surveys also report increasing use of LLMs across software engineering tasks~\cite{fan2023large,hou2024large,zhang2026survey}. These studies demonstrate that LLMs can generate useful requirements artifacts, whereas our work studies a different problem: how legacy SRS content can be decomposed into traceable atomic evidence, routed to typed pre-SRS artifacts, and evaluated for downstream reconstruction.

NLP4RE has long studied requirements classification, extraction, ambiguity analysis, traceability, and quality assessment~\cite{zhao2021natural,abualhaija2024replication,hey2020norbert,dalpiaz2019requirements,cleland2007automated,genova2013framework}. Recent work also explores extracting models or requirements from agile backlogs, app reviews, and other textual sources~\cite{10.1109/MODELS-C59198.2023.00096,10.1145/3412841.3442006}, as well as using NLI for RE classification and defect or conflict analysis~\cite{fazelnia2024lessons}. Closest to our setting, Okamoto et al.~\cite{okamoto2025restructuring} study LLM-based restructuring of organically written SRSs into ISO/IEC/IEEE 29148-compliant documents. Our work shares the goal of making legacy SRSs more structured and analyzable, but operates at a finer granularity by decomposing sections into atomic propositions before routing them to pre-SRS artifact types.

Public datasets such as PURE~\cite{ferrari2017pure} support reproducible NLP4RE research, while recent work calls for reusable benchmarks, clearer reporting, and stronger traceability resources~\cite{abualhaija2024replication,10.1145/3786771}. Traceability recovery research further emphasizes linking information across heterogeneous lifecycle artifacts~\cite{gao2024triad}, and synthetic requirements data has been explored as a way to address data scarcity in AI4RE~\cite{el2025good}. \name contributes to this direction by using public SRSs as seed evidence for constructing traceable synthetic pre-SRS artifacts.

\section{Threats to Validity}
\label{sec:threats}

\textbf{Construct validity.}
Our metrics measure traceability-grounded proxy quality, not the complete correctness of requirements. AlignScore approximates factual support, SBERT cosine semantic recoverability, and judge/Prometheus scores rubric-based quality; none establishes full SRS coherence, ISO conformance, or practitioner usefulness. Because PURE and PROMISE-NFR are public, strong scores may partly reflect benchmark exposure rather than only decomposition or routing ability \cite{riddell2024quantifying}. We mitigate this through stage-wise evaluation under our routing schema and triangulation across evidence on grounding, coverage, quality, agreement, revision, failure, and retention.

\textbf{Internal validity.}
The pipeline contains several LLM-based stages, so outputs may vary with prompts, model updates, decoding, and nondeterminism. We reduce this risk using fixed prompts, codebooks, traceability identifiers, plurality voting, judge-guided refinement, and manual input checks. The temperature was set to 1.0 to encourage diversity, but single-run outputs cannot establish model stability.

\textbf{External validity.}
We evaluate seven selected text-based PURE SRS documents and synthetic pre-SRS artifact types; results may differ for proprietary, safety-critical, table-heavy, multimodal, or actively maintained industrial requirements. We mitigate this through multiple SRSs, models, artifact types, and released prompts, codebooks, artifacts, and outputs. We did not run a baseline comparison because the study evaluates LLM behavior in a controlled traceability setting, not \name as a competing SRS-generation method; existing baselines do not expose the intermediate provenance needed for our analyses.

\textbf{Conclusion validity.}
Several analyses use a few SRS documents, nested units, fixed thresholds, and one generation run per model. We mitigate this through dataset-level aggregation, distributional summaries, paired bootstrap comparisons where applicable, and cautious interpretation. Overall, our conclusions support feasibility, observed tradeoffs, and failure-mode characterization, not statistical superiority, production readiness, or deployment-level effectiveness.

\section{Conclusion}
\label{sec:conclusion}
We conducted an empirical study of how LLMs transform legacy SRS documents into traceable synthetic pre-SRS artifacts, using \name as a controlled study pipeline. 
Across seven selected text-only PURE SRSs, our results show that LLMs can produce faithful and usable atomic statements and grounded typed artifacts, but these transformations are not lossless. Semantic routing and quality judgment require human-grounded model selection, artifact grounding does not guarantee structural completeness, and downstream reconstruction preserves source information only partially. These findings suggest that traceable synthetic artifacts can support fine-grained evaluation of LLM-based SRS generation, while exposing important tradeoffs among faithfulness, coverage, sparsity, and recoverability. Future work should extend this study to multimodal and industrial SRSs, evaluate the usefulness of artifacts with RE practitioners, and develop loss-aware decomposition and more flexible artifact schemas for partially specified source evidence.

\section{Data Availability}
\label{sec:datapackage}
To support double-anonymous review and reproducibility, we provide an anonymized stable repository~\cite{ReqGenXZenodo}. The repository includes processed inputs, generated atoms, semantic tags, routed pre-SRS artifacts, reconstructed SRS outputs, traceability metadata, evaluation outputs, scripts, notebooks, prompts, rubrics/codebooks, templates, configuration files, cached outputs, and README files. During review, author-identifying metadata, commit history, acknowledgments, and ownership information are removed or hidden. Upon acceptance, we will archive the final package with a permanent DOI.


\bibliography{biblography}

@inproceedings{reimers2019sentence,
  title={Sentence-bert: Sentence embeddings using siamese bert-networks},
  author={Reimers, Nils and Gurevych, Iryna},
  booktitle={Proceedings of the 2019 conference on empirical methods in natural language processing and the 9th international joint conference on natural language processing (EMNLP-IJCNLP)},
  pages={3982--3992},
  year={2019}
}

@article{zhang2026survey,
  title={A survey on large language models for software engineering},
  author={Zhang, Quanjun and Fang, Chunrong and Xie, Yang and Zhang, Yaxin and Yu, Shengcheng and Sun, Weisong and Yang, Yun and Chen, Zhenyu},
  journal={Science China Information Sciences},
  volume={69},
  number={4},
  pages={141102},
  year={2026},
  publisher={Springer}
}

@article{hou2024large,
  title={Large language models for software engineering: A systematic literature review},
  author={Hou, Xinyi and Zhao, Yanjie and Liu, Yue and Yang, Zhou and Wang, Kailong and Li, Li and Luo, Xiapu and Lo, David and Grundy, John and Wang, Haoyu},
  journal={ACM Transactions on Software Engineering and Methodology},
  volume={33},
  number={8},
  pages={1--79},
  year={2024},
  publisher={ACM New York, NY}
}

@article{quattrocchi2026can,
  title={Can llms generate user stories and assess their quality?},
  author={Quattrocchi, Giovanni and Pasquale, Liliana and Spoletini, Paola and Baresi, Luciano},
  journal={IEEE Transactions on Software Engineering},
  year={2026},
  publisher={IEEE}
}

@INPROCEEDINGS {okamoto2025restructuring,
author = { Okamoto, Ryu and Kusumoto, Shinji },
booktitle = { 2025 IEEE 33rd International Requirements Engineering Conference (RE) },
title = {{ Towards the Automatic Restructuring of Software Requirements Specifications to Conform to Standards Using Large Language Models }},
year = {2025},
volume = {},
ISSN = {},
pages = {467-475},
doi = {10.1109/RE63999.2025.00056},
url = {https://doi.ieeecomputersociety.org/10.1109/RE63999.2025.00056},
publisher = {IEEE Computer Society},
address = {Los Alamitos, CA, USA},
month =sep}

@inproceedings{zhu2025conformity,
  title={Conformity in large language models},
  author={Zhu, Xiaochen and Zhang, Caiqi and Stafford, Tom and Collier, Nigel and Vlachos, Andreas},
  booktitle={Proceedings of the 63rd Annual Meeting of the Association for Computational Linguistics (Volume 1: Long Papers)},
  pages={3854--3872},
  year={2025}
}

@article{wang2022self,
  title={Self-consistency improves chain of thought reasoning in language models},
  author={Wang, Xuezhi and Wei, Jason and Schuurmans, Dale and Le, Quoc and Chi, Ed and Narang, Sharan and Chowdhery, Aakanksha and Zhou, Denny},
  journal={arXiv preprint arXiv:2203.11171},
  year={2022}
}

@inproceedings{hosseini2024scalable,
  title={Scalable and domain-general abstractive proposition segmentation},
  author={Hosseini, Mohammad Javad and Gao, Yang and Baumg{\"a}rtner, Tim and Fabrikant, Alex and Amplayo, Reinald Kim},
  booktitle={Findings of the Association for Computational Linguistics: EMNLP 2024},
  pages={8856--8872},
  year={2024}
}

@inproceedings{chen2024dense,
  title={Dense x retrieval: What retrieval granularity should we use?},
  author={Chen, Tong and Wang, Hongwei and Chen, Sihao and Yu, Wenhao and Ma, Kaixin and Zhao, Xinran and Zhang, Hongming and Yu, Dong},
  booktitle={Proceedings of the 2024 Conference on Empirical Methods in Natural Language Processing},
  pages={15159--15177},
  year={2024}
}

@misc{ReqGenXZenodo,
  author       = {{Anonymous}},
  title        = {{ReqGenX}},
  howpublished = {\url{https://doi.org/10.5281/zenodo.20267334}},
  publisher    = {Zenodo},
  year         = {2026},
  doi          = {10.5281/zenodo.20267334},
}

@inproceedings{hey2020norbert,
  title={Norbert: Transfer learning for requirements classification},
  author={Hey, Tobias and Keim, Jan and Koziolek, Anne and Tichy, Walter F},
  booktitle={2020 IEEE 28th international requirements engineering conference (RE)},
  pages={169--179},
  year={2020},
  organization={IEEE}
}

@inproceedings{ronanki2023investigating,
  title={Investigating chatgpt’s potential to assist in requirements elicitation processes},
  author={Ronanki, Krishna and Berger, Christian and Horkoff, Jennifer},
  booktitle={2023 49th Euromicro conference on software engineering and advanced applications (SEAA)},
  pages={354--361},
  year={2023},
  organization={IEEE}
}

@article{genova2013framework,
  title={A framework to measure and improve the quality of textual requirements},
  author={G{\'e}nova, Gonzalo and Fuentes, Jos{\'e} M and Llorens, Juan and Hurtado, Omar and Moreno, Valent{\'\i}n},
  journal={Requirements engineering},
  volume={18},
  number={1},
  pages={25--41},
  year={2013},
  publisher={Springer}
}

@inproceedings{bacchelli2011extracting,
  title={Extracting structured data from natural language documents with island parsing},
  author={Bacchelli, Alberto and Cleve, Anthony and Lanza, Michele and Mocci, Andrea},
  booktitle={2011 26th IEEE/ACM International Conference on Automated Software Engineering (ASE 2011)},
  pages={476--479},
  year={2011},
  organization={IEEE}
}

@inproceedings{gao2024triad,
  title={Triad: Automated traceability recovery based on biterm-enhanced deduction of transitive links among artifacts},
  author={Gao, Hui and Kuang, Hongyu and Assun{\c{c}}{\~a}o, Wesley KG and Mayr-Dorn, Christoph and Rong, Guoping and Zhang, He and Ma, Xiaoxing and Egyed, Alexander},
  booktitle={Proceedings of the IEEE/ACM 46th International Conference on Software Engineering},
  pages={1--13},
  year={2024}
}

@inproceedings{fan2023large,
  title={Large language models for software engineering: Survey and open problems},
  author={Fan, Angela and Gokkaya, Beliz and Harman, Mark and Lyubarskiy, Mitya and Sengupta, Shubho and Yoo, Shin and Zhang, Jie M},
  booktitle={2023 IEEE/ACM International Conference on Software Engineering: Future of Software Engineering (ICSE-FoSE)},
  pages={31--53},
  year={2023},
  organization={IEEE}
}

@inproceedings{zhu2025reqinone,
  title={ReqInOne: A Large Language Model-Based Agent for Software Requirements Specification Generation},
  author={Zhu, Taohong and Cordeiro, Lucas C and Sun, Youcheng},
  booktitle={2025 IEEE 33rd International Requirements Engineering Conference (RE)},
  pages={449--457},
  year={2025},
  organization={IEEE}
}

@inproceedings{fazelnia2024lessons,
  title={Lessons from the use of natural language inference (nli) in requirements engineering tasks},
  author={Fazelnia, Mohamad and Koscinski, Viktoria and Herzog, Spencer and Mirakhorli, Mehdi},
  booktitle={2024 IEEE 32nd International Requirements Engineering Conference (RE)},
  pages={103--115},
  year={2024},
  organization={IEEE}
}

@inproceedings{dalpiaz2019requirements,
  title={Requirements classification with interpretable machine learning and dependency parsing},
  author={Dalpiaz, Fabiano and Dell'Anna, Davide and Aydemir, Fatma Basak and {\c{C}}evikol, Sercan},
  booktitle={2019 IEEE 27th International Requirements Engineering Conference (RE)},
  pages={142--152},
  year={2019},
  organization={IEEE}
}

@article{cleland2007automated,
  title={Automated classification of non-functional requirements},
  author={Cleland-Huang, Jane and Settimi, Raffaella and Zou, Xuchang and Solc, Peter},
  journal={Requirements engineering},
  volume={12},
  number={2},
  pages={103--120},
  year={2007},
  publisher={Springer}
}

@inproceedings{ferrari2017pure,
  title={Pure: A dataset of public requirements documents},
  author={Ferrari, Alessio and Spagnolo, Giorgio Oronzo and Gnesi, Stefania},
  booktitle={2017 IEEE 25th international requirements engineering conference (RE)},
  pages={502--505},
  year={2017},
  organization={IEEE}
}

@inproceedings{zha2023alignscore,
  title={AlignScore: Evaluating factual consistency with a unified alignment function},
  author={Zha, Yuheng and Yang, Yichi and Li, Ruichen and Hu, Zhiting},
  booktitle={Proceedings of the 61st Annual Meeting of the Association for Computational Linguistics (Volume 1: Long Papers)},
  pages={11328--11348},
  year={2023}
}

@inproceedings{krishna2024using,
  title={Using llms in software requirements specifications: An empirical evaluation},
  author={Krishna, Madhava and Gaur, Bhagesh and Verma, Arsh and Jalote, Pankaj},
  booktitle={2024 IEEE 32nd International Requirements Engineering Conference (RE)},
  pages={475--483},
  year={2024},
  organization={IEEE}
}

@inproceedings{pasquale2025exploring,
  title={Exploring the use of llms for requirements specification in an it consulting company},
  author={Pasquale, Liliana and Ragone, Azzurra and Piemontese, Emanuele and Darban, Armin Amiri},
  booktitle={2025 IEEE 33rd International Requirements Engineering Conference (RE)},
  pages={389--399},
  year={2025},
  organization={IEEE}
}

@inproceedings{lazik2025good,
  title={The Good, the Bad, and the Uncanny: Investigating Diversity Aspects of LLM-Generated Personas for Requirements Engineering},
  author={Lazik, Christopher and Kauter, Charlotte and Nunes, In{\^e}s and Ziglowski, Aaron and Pryma, Alina and Katins, Christopher and Grunske, Lars and Kosch, Thomas},
  booktitle={2025 IEEE 33rd International Requirements Engineering Conference (RE)},
  pages={244--256},
  year={2025},
  organization={IEEE}
}

@article{el2025good,
  title={How good are synthetic requirements? Evaluating LLM-Generated datasets for AI4RE},
  author={El-Hajjami, Abdelkarim and Salinesi, Camille},
  journal={arXiv preprint arXiv:2506.21138},
  year={2025}
}

@inproceedings{xie2025effective,
  title={How effective are large language models in generating software specifications?},
  author={Xie, Danning and Yoo, Byoungwoo and Jiang, Nan and Kim, Mijung and Tan, Lin and Zhang, Xiangyu and Lee, Judy S},
  booktitle={2025 IEEE International Conference on Software Analysis, Evolution and Reengineering (SANER)},
  pages={1--12},
  year={2025},
  organization={IEEE}
}

@inproceedings{spijkman2025llm,
  title={LLM-Assisted Requirements Engineering in Agile MDD: Industry Insights and Validation},
  author={Spijkman, Tjerk and Molenkamp, Bente and Beudeker, Steffen and Overbeek, Sietse and Dalpiaz, Fabiano},
  booktitle={2025 IEEE 33rd International Requirements Engineering Conference (RE)},
  pages={366--377},
  year={2025},
  organization={IEEE}
}

@article{zhao2021natural,
  title={Natural language processing for requirements engineering: A systematic mapping study},
  author={Zhao, Liping and Alhoshan, Waad and Ferrari, Alessio and Letsholo, Keletso J and Ajagbe, Muideen A and Chioasca, Erol-Valeriu and Batista-Navarro, Riza T},
  journal={ACM Computing Surveys (CSUR)},
  volume={54},
  number={3},
  pages={1--41},
  year={2021},
  publisher={ACM New York, NY, USA}
}

@article{shirabad2005promise,
  title={The PROMISE repository of software engineering databases},
  author={Shirabad, SAYYAD},
  journal={(No Title)},
  year={2005}
}

@dataset{li2024desiree,
  author       = {Li, F.-L. and Horkoff, J. and Mylopoulos, J. and Borgida, A. and Guizzardi, R. and Guizzardi, G. and Liu, L. and Peng, Y.},
  title        = {Dataset for a requirements modeling language and the Desiree framework},
  year         = {2024},
  publisher    = {Zenodo},
  doi          = {10.5281/zenodo.14330485},
  url          = {https://doi.org/10.5281/zenodo.14330485},
  note         = {Data set}
}

@ARTICLE{8559686,
  author={},
  journal={ISO/IEC/IEEE 29148:2018(E)}, 
  title={{ISO/IEC/IEEE} International Standard - Systems and software engineering -- Life cycle processes -- Requirements engineering}, 
  year={2018},
  volume={},
  number={},
  pages={1-104},
  doi={10.1109/IEEESTD.2018.8559686}}

@article{cockburn1998basic,
  title={Basic use case template},
  author={Cockburn, Alistair},
  journal={Humans and Technology, Technical Report},
  volume={96},
  pages={28},
  year={1998}
}

@article{robertson2000volere,
  title={Volere},
  author={Robertson, James and Robertson, Suzanne},
  journal={Requirements Specification Templates},
  year={2000}
}

@book{brennan2009guide,
  title={A guide to the Business Analysis Body of Knowledge (BABOK Guide)},
  author={Brennan, Kevin and others},
  year={2009},
  publisher={IIBA}
}

@book{starke2019arc42,
  title={arc42 by Example},
  author={Starke, Gernot and Simons, Michael and Zorner, Stefan and M{\"u}ller, Ralf},
  year={2019},
  publisher={Packt Publishing}
}

@ARTICLE{392555,
  author={},
  journal={IEEE Std 830-1993}, 
  title={{IEEE} Recommended Practice for Software Requirements Specifications}, 
  year={1994},
  volume={},
  number={},
  pages={1-32},
  doi={10.1109/IEEESTD.1994.121431}}

@inproceedings{kim2024prometheus,
  title={Prometheus 2: An open source language model specialized in evaluating other language models},
  author={Kim, Seungone and Suk, Juyoung and Longpre, Shayne and Lin, Bill Yuchen and Shin, Jamin and Welleck, Sean and Neubig, Graham and Lee, Moontae and Lee, Kyungjae and Seo, Minjoon},
  booktitle={Proceedings of the 2024 Conference on Empirical Methods in Natural Language Processing},
  pages={4334--4353},
  year={2024}
}

@inproceedings{10.1145/3412841.3442006,
author = {de Ara\'{u}jo, Adailton Ferreira and Marcacini, Ricardo Marcondes},
title = {RE-BERT: automatic extraction of software requirements from app reviews using BERT language model},
year = {2021},
isbn = {9781450381048},
publisher = {Association for Computing Machinery},
address = {New York, NY, USA},
url = {https://doi.org/10.1145/3412841.3442006},
doi = {10.1145/3412841.3442006},
booktitle = {Proceedings of the 36th Annual ACM Symposium on Applied Computing},
pages = {1321–1327},
numpages = {7},
location = {Virtual Event, Republic of Korea},
series = {SAC '21}
}

@article{10.1145/3786771,
author = {Hu, Xing and Niu, Feifei and Chen, Junkai and Zhou, Xin and Zhang, Junwei and He, Junda and Xia, Xin and Lo, David},
title = {Assessing and Advancing Benchmarks for Evaluating Large Language Models in Software Engineering Tasks},
year = {2026},
publisher = {Association for Computing Machinery},
address = {New York, NY, USA},
issn = {1049-331X},
url = {https://doi.org/10.1145/3786771},
doi = {10.1145/3786771},
note = {Just Accepted},
journal = {ACM Trans. Softw. Eng. Methodol.},
month = dec
}

@inproceedings{10.1109/MODELS-C59198.2023.00096,
author = {Arulmohan, Sathurshan and Meurs, Marie-Jean and Mosser, S\'{e}bastien},
title = {Extracting Domain Models from Textual Requirements in the Era of Large Language Models},
year = {2023},
publisher = {IEEE Press},
url = {https://doi.org/10.1109/MODELS-C59198.2023.00096},
doi = {10.1109/MODELS-C59198.2023.00096},
booktitle = {2023 ACM/IEEE International Conference on Model Driven Engineering Languages and Systems Companion (MODELS-C)},
pages = {580–587},
numpages = {8},
location = {V\"{a}ster\r{a}s, Sweden}
}

@article{abualhaija2024replication,
author = {Abualhaija, Sallam and Aydemir, F. Basak and Dalpiaz, Fabiano and Dell'Anna, Davide and Ferrari, Alessio and Franch, Xavier and Fucci, Davide},
title = {Replication in Requirements Engineering: The NLP for RE Case},
year = {2024},
issue_date = {July 2024},
publisher = {Association for Computing Machinery},
address = {New York, NY, USA},
volume = {33},
number = {6},
issn = {1049-331X},
url = {https://doi.org/10.1145/3658669},
doi = {10.1145/3658669},
journal = {ACM Trans. Softw. Eng. Methodol.},
month = jun,
articleno = {151},
numpages = {33}
}

@inproceedings{riddell2024quantifying,
  title={Quantifying contamination in evaluating code generation capabilities of language models},
  author={Riddell, Martin and Ni, Ansong and Cohan, Arman},
  booktitle={Proceedings of the 62nd Annual Meeting of the Association for Computational Linguistics (Volume 1: Long Papers)},
  pages={14116--14137},
  year={2024}
}

\end{document}